\documentclass[article, twocolumn, 
   aps, pra,
  amsmath,amssymb,
  longbibliography,
  ]{revtex4-1}
\usepackage{graphicx,color}
\usepackage{amsmath}
\usepackage{natbib}
\usepackage{epsfig}
\begin{document}

\title{\color{blue} Excess entropy determines the applicability of Stokes-Einstein relation in simple fluids}

\author{S. A. Khrapak}\email{Sergey.Khrapak@gmx.de}
\affiliation{Joint Institute for High Temperatures, Russian Academy of Sciences, 125412 Moscow, Russia}
\author{A. G. Khrapak}
\affiliation{Joint Institute for High Temperatures, Russian Academy of Sciences, 125412 Moscow, Russia}

\begin{abstract}
The Stokes-Einstein (SE) relation between the self-diffusion and shear viscosity coefficients operates in sufficiently dense liquids not too far from the liquid-solid phase transition. By considering four simple model systems with very different pairwise interaction potentials (Lennard-Jones, Coulomb, Debye-H\"uckel or screened Coulomb, and the hard sphere limit) we identify where exactly on the respective phase diagrams the SE relation holds. It appears that the reduced excess entropy $s_{\rm ex}$ can be used as a suitable  indicator of the validity of the SE relation. In all cases considered the onset of SE relation validity occurs at approximately $s_{\rm ex}\lesssim -2$. In addition, we demonstrate that the line separating gas-like and liquid-like fluid behaviours on the phase diagram is roughly characterized by $s_{\rm ex}\simeq -1$. 
\end{abstract}

\date{\today}

\maketitle

\section{Introduction}

An accurate general theory of transport process in liquids is still lacking, despite considerable progress achieved over many decades~\cite{FrenkelBook,HansenBook,GrootBook,MarchBook}. 
Moreover, it is very unlikely that a general theory of transport processes in liquids can be constructed at all.
Difficulties with theoretical description of liquid state dynamics in comparison with solids and gases can be understood as follows~\cite{BrazhkinUFN2017}. Solids and gases can be considered in some (dynamical) sense as ``pure'' aggregate states. In solids the motion of atoms is purely vibrational, while in dilute gases atoms move freely along straight trajectories between collisions. This simplifies the development of transport theories. In this context liquids constitute a ``mixed'' aggregate state. Both vibrational and ballistic atomic motions are present. Their relative importance depends on the location on the phase diagram. Near the liquid-solid phase transition vibrational motion dominates and solid-like approaches to transport properties are more relevant. At lower densities and higher temperatures ballistic motion is more important and transport is similar to that in dense non-ideal gases.      

In the absence of general theories we often have to rely (when experimental data is not available) on phenomenological approaches, semi-quantitative models, and scaling relationships. In this context the Stokes-Einstein relation between the self-diffusion and shear viscosity coefficients has been proven to be particularly useful. 

The conventional Stokes-Einstein (SE) relation expresses the diffusion coefficient $D$ of a tracer macroscopic spherical (``Brownian'') particle of radius $R$ in terms of the temperature $T$ (expressed in energy units, $k_{\rm B}=1$) and shear viscosity  coefficient $\eta$ of a medium it is immersed in. It reads~\cite{BalucaniBook}
\begin{equation}\label{SE_1}
D=\frac{T}{c\pi \eta R},
\end{equation}   
where $c$ is a numerical coefficient: $c=6$ or $c=4$ corresponds to the ``stick'' or ``slip'' boundary condition at the sphere surface, respectively. When the size of the tracer sphere decreases, the actual size of the sphere in Eq.~(\ref{SE_1}) has to be replaced by the so-called hydrodynamic radius $R_{\rm H}$, which can depend on details of the interaction between the sphere and the atoms or molecules of the medium. Going further down to atomistic scales, when self-diffusion of atoms in simple pure fluids is considered, the SE relation takes the form\begin{equation}\label{SE_2}
D\eta(\Delta/T)=\alpha_{\rm SE},
\end{equation}
where $\Delta=\rho^{-1/3}$ is the mean interatomic separation, which now plays the role of the effective tracer sphere diameter, and $\rho$ is the atomic number density. 
Equation (\ref{SE_2}) is also known as the Stokes-Einstein relation without the hydrodynamic diameter~\cite{CostigliolaJCP2019}.
In view of Eq.~(\ref{SE_1}) the coefficient $\alpha_{\rm SE}$ can be expected to vary between $1/3\pi\simeq 0.106$ and $1/2\pi\simeq 0.159$ for the stick and slip boundary condition, respectively. 

The relation of the form of Eq.~(\ref{SE_2}) appeared already in the book by Frenkel~\cite{FrenkelBook} when he discussed the viscosity of simple liquids. He also provided some qualitative arguments regarding why a macroscopic approach can be applied down to atomic scales. At the same time, he pointed out that the derivation used to arrive at Eq.~(\ref{SE_2}) is rather formal and does not explain the exact mechanism that determines the viscosity coefficient.

The purpose of the present paper is to provide a systematic consistent picture concerning the applicability of the SE relation to simple fluids. We analyse contemporary transport data for four different model system: Lennard-Jones liquids, one-component plasma, Yukawa and hard-sphere fluids in order to identify how general is the SE relation. Three main questions are to be answered: (i) Does SE relation applies to each concrete equilibrium fluid? (ii) What is the value of the SE coefficient $\alpha_{\rm SE}$ and are there correlations with the properties of interparticle interaction? (iii) Where is the regime of SE relation applicability on the phase diagram and whether there exists a universal criterion of its applicability? These questions will be addressed in the following sections. A very important question related to the breakdown of the SE relation in supercooled and glass forming liquids~\cite{HodgdonPRE1993,TarjusJCP1995,BordatJPCM2003,ChenPNAS2006,PuosiJCP2018}    
is beyond the scope of this paper.

\section{Theoretical background}          

\subsection{Excess entropy scaling (Rosenfeld)}

In 1977 Rosenfeld proposed a relation beween transport coefficients and internal entropy of simple systems~\cite{RosenfeldPRA1977}. In particular, he demonstrated that properly reduced diffusion and shear viscosity coefficients are approximately exponential functions of the reduced excess entropy $s_{\rm ex}=(S-S_{\rm id})/Nk_{\rm B}$, where $S$ is the system entropy, $S_{\rm id}$ is the entropy of the ideal gas at the same temperature and density, $N$ is the number of particles and $k_{\rm B}$ is the  Boltzmann's constant.
The system-independent normalization for the transport coefficients used by Rosenfeld reads:
\begin{equation}\label{Rosenfeld}
D_{\rm R}  =  D\frac{\rho^{1/3}}{v_{\rm T}} , \quad\quad
\eta_{\rm R}  =  \eta \frac{\rho^{-2/3}}{m v_{\rm T}},  
\end{equation}
where $v_{\rm T}=\sqrt{T/m}$ is the thermal velocity and $m$ is the atomic mass. A somewhat different variant of entropy scaling of atomic diffusion in condensed matter was also proposed by Dzugutov~\cite{DzugutovNature1996}, who used the excess entropy in the pair approximation instead of the full excess entropy (note that the total excess entropy can be approximated by the pair contribution only in some vicinity of the freezing point~\cite{LairdPRA1992,GiaquintaPhysA1992,GiaquintaPRA1992,
SaijaJCP2006,FominJCP2014,KlumovResPhys2020,KhrapakJCP2021}). By now it is well recognized that many simple and not so simple systems conform to the approximate excess entropy scaling. There are also counterexamples, where the original excess entropy scaling is not applicable~\cite{KrekelbergPRE03_2009,KrekelbergPRE12_2009,
FominPRE2010}. For a recent review of this topic see e.g. Ref.~\cite{DyreJCP2018}.

For our present purpose we quote the approximate formulas for the diffusion and viscosity coefficients proposed by Rosenfeld~\cite{RosenfeldJPCM1999}   
\begin{equation}\label{Rosenfeld1}
D_{\rm R}  \simeq  0.6{\rm e}^{0.8 s_{\rm ex}} , \quad\quad
\eta_{\rm R}  =  0.2{\rm e}^{-0.8 s_{\rm ex}}.  
\end{equation}
Combining this with the normalization properties of Eq.~(\ref{Rosenfeld}) we immediately arrive at
\begin{equation}
D_{\rm R}\eta_{\rm R}\equiv D\eta\left(\frac{\Delta}{T}\right)\simeq 0.12.
\end{equation}
Thus, the SE relation of the form (\ref{SE_2}) is automatically satisfied with $\alpha_{\rm SE}\simeq 0.12$, which is relatively close to real SE coefficients for sufficiently soft interactions (see below). However, this observation is merely heuristic and does not explain physical mechanisms relating viscosity and diffusion.   

\subsection{The velocity field approach}

A microscopic velocity field approach was proposed by Gaskell and Miller~\cite{GaskellJPC1978} to include microscopic features into hydrodynamic description. It allows to express the velocity autocorrelation function in terms of the longitudinal and transverse current correlation functions. The result is similar to that of the mode coupling theory, the only difference is the presence of a form factor under the integral. The Green-Kubo formula is then used to obtain the self-diffusion coefficient. The contribution from the longitudinal correlations cancels out exactly and the diffusion coefficient is related to the properties of the transverse mode. The derivation is rather involved and is not presented here, although no principal difficulties arise. We quote the final result~\cite{BalucaniBook,GaskellPLA1982,BalucaniJPC1985,Balucani1990} 
\begin{equation}
D=\frac{T}{4\pi \eta a},
\end{equation}
where $a=(4\pi\rho/3)^{-1/3}$ is the Wigner-Seitz radius.
The emerging SE coefficient $\alpha_{\rm SE}= (\Delta/a)/4\pi\simeq 0.13$ is appropriate (see below). The derivation tells us that the relation between the diffusion and shear viscosity coefficients comes from the properties of the transverse collective mode. It can be made more transparent by taking the hydrodynamic limit for the transverse current correlation function, namely $C_T (q, t) = (T/m)\exp[ -(\eta/\rho m) q^2 t]$, where $q$ is the wave vector~\cite{Balucani1990}. Still no simple quantitative picture regarding the physical mechanisms behind the SE relation immediately emerges.    

\subsection{Damped oscillator model (Zwanzig)}
 
Perhaps one of the simplest and transparent variants of the derivation of SE relations for simple fluids was presented by Zwanzig based on the relations between the transport coefficients and properties of collective excitations~\cite{ZwanzigJCP1983}. 

Zwanzig's approach is based on the assumption that atoms in liquids exhibit solid-like oscillations about temporary equilibrium positions corresponding to a local minimum on the system's potential energy surface~\cite{FrenkelBook,Stillinger1982}. These positions do not form a regular lattice like in crystalline solids. They are also not fixed and change with time (this is why liquids can flow). It can be assumed that a local configuration is preserved for some time until a fluctuation in the kinetic energy allows to rearrange the positions of some of the atoms towards a new local minimum in the potential energy surface. The magnitude of these rearrangements (''cell jumps'' in Zwanzig's terminology) is irrelevant for the present consideration. The waiting time distribution of the rearrangements scales exponentially, $\propto\exp(-t/\tau)$,  where $\tau$ is a lifetime. Atomic motions after the rearrangements are uncorrelated with motions before rearrangements. The lifetime $\tau$ should be considerably longer than the characteristic period of solid-like vibrations for the dynamical picture sketched makes sense.

Within this ansatz a reasonable approximation for the velocity autocorrelation function of an atom $j$ is 
\begin{equation}
Z_j(t)\simeq \left(\frac{T}{m}\right)\cos (\omega_j t)\exp(-t/\tau),
\end{equation}
corresponding to a time dependence of a damped harmonic oscillator. The self-diffusion coefficient $D$ is given by the Green-Kubo formula
\begin{equation}
D=\frac{1}{N}\int_0^{\infty}\sum_j Z_j(t)dt.
\end{equation}
Zwanzig then assumed that vibrational frequencies $\omega_j$ are related to the collective mode spectrum and performed averaging over collective modes. Since the exact distribution of frequencies is generally not available, he used a Debye approximation, characterized by one longitudinal and two transverse modes with acoustic dispersion. The result is
\begin{equation}\label{DZ}
D=\frac{T}{3\pi}\left(\frac{3\rho}{4\pi}\right)^{1/3}\left(\frac{1}{\rho mc_l^2\tau}+\frac{2}{\rho mc_t^2\tau}\right),
\end{equation} 
where $c_l$ and $c_t$ are the (instantaneous) longitudinal and transverse sound velocities, related to the elastic response of fluids to high-frequency perturbations~\cite{ZwanzigJCP1965}. 
The last step is to set the lifetime $\tau$ equal to the Maxwellian shear relaxation time~\cite{KhrapakMolPhys2019},  
\begin{equation}
\tau= {\eta}/G_{\infty}, 
\end{equation}
where $G_{\infty}=\rho m c_t^2$ is the infinite frequency (instantaneous) shear modulus. The idea that the residence time of atoms in their temporary equlibrium positions should be associated with Maxwellian relaxation time was discussed already by Frenkel~\cite{FrenkelBook}. With this we obtain the SE relation of the form
\begin{equation}\label{DZ1}
D\eta(\Delta/T)\equiv\alpha_{\rm SE}\simeq 0.13\left(1+\frac{c_t^2}{2c_l^2}\right).
\end{equation}
Zwanzig did not explicitly assumed that $\tau$ is given be the Maxwellian relaxation time and expressed the SE coefficient in terms of the longitudinal and shear viscosities $\alpha_{\rm SE}\simeq 0.13 (1+\eta/2\eta_{l})$. Otherwise, Eq.~(\ref{DZ1}) is identical to the original result. Note that since the sound velocity ratio $c_t/c_l$ is confined in the range from $0$ to $\sqrt{3}/2$, the coefficient $\alpha_{\rm SE}$ can vary
only between $\simeq 0.13$ and $\simeq 0.18$~\cite{ZwanzigJCP1983,KhrapakMolPhys2019}. 

Zwanzig noted that ``all of these assumptions are arguable, though plausible as a first guess, and could be tested by deeper theoretical analysis and by molecular dynamics simulations''. Some of the assumptions, in particular the effect of the waiting time distribution were critically checked in Ref.~\cite{MohantyPRA1985}. The waiting times were also estimated using MD simulations by introducing the cage correlation functions which measure the rate of change of atomic surroundings~\cite{RabaniJCP1997}. 

Nevertheless, despite of the simplifications involved the predictive power of Zwangig's model is impressive. Eq.~(\ref{SE_2}) is satisfied to a very high accuracy in some vicinity of the liquid-solid phase transition of many simple model liquids~\cite{CostigliolaJCP2019,KhrapakMolPhys2019}. Moreover, the coefficient $\alpha_{\rm SE}$ can be correlated with the potential softness (via the ratio of the longitudinal and transverse sound velocities), as the model predicts~\cite{KhrapakMolPhys2019}. For soft long-ranged interactions (such as e.g. Coulomb or screened Coulomb) the strong inequality $c_l\gg c_t$ is satisfied and $\alpha_{\rm SE}$ tends to its lower limit. For steeper potentials (such as e.g. Lennard-Jones) the ratio $c_t/c_l$ increases and $\alpha_{\rm SE}$ increases too, as expected. For many liquid metals at the melting temperature the coefficient $\alpha_{\rm SE}$ is located in the vicinity of $\simeq 0.15$, although considerably higher values were also reported~\cite{KhrapakMolPhys2019}. 

According to the Zwanzig's model the SE relation is applicable for liquids close to the melting temperatures and densities, but not at high temperatures and low densities. The dynamical picture involves fast solid-like oscillations around temporary equilibrium positions. This is clearly irrelevant for gases. In dilute gases the atoms move freely between collisions. For a dilute gas of hard spheres (HS) of diameter $\sigma$ the self-diffusion and viscosity coefficients are given in the first approximation by~\cite{LifshitzKinetics}  
\begin{equation}
D = \frac{3}{8\rho\sigma^2}\left(\frac{T}{\pi m}\right)^{1/2}, \quad\quad \eta = \frac{5}{16\sigma^2}\left(\frac{mT}{\pi}\right)^{1/2}. \label{HStransport}
\end{equation}
Thus the relation between the diffusion and viscosity coefficients in the gaseous phase is $\eta\sim m\rho D$ and the relation (\ref{SE_2}) cannot be satisfied. This result is quite general and does not depend on the specifics of the HS system (according to elementary gas-kinetic formulas $D\sim v\ell$ and $\eta\sim m \rho v \ell$, where $v$ is the mean atomic velocity and $\ell$ is the mean free path between atomic collisions~\cite{LifshitzKinetics}).   

The relevant question is therefore how far from the freezing line the SE relation can operate and what its applicability conditions are. From the dynamical picture sketched above it is obvious that the condition $\omega\tau >1$ should be at least satisfied. A characteristic vibrational frequency $\omega$ can be associated with the Einstein frequency $\Omega_{\rm E}$. Still, we would need to know the behaviour of the shear viscosity $\eta$ and shear modulus $G_{\infty}$ in order to evaluate $\tau$ and estimate $\tau\Omega_{\rm E}$. This is not very realistic in general (although for some special systems this program is feasible). Other options should be considered.     

Below we present the evidence based on the analysis of extensive simulation data that there is a relatively wide region prior to freezing where the SE relation holds.
The onset of its validity can be conveniently characterized by the magnitude of the excess entropy. Quantitatively, the SE relation in the liquid phase is valid for $s_{\rm ex}\lesssim -2$. Our results can also shed some light on the dynamical crossover separating the liquid-like from a gas-like regions on a phase diagram (the so-called ``Frenkel line''). This is the topic of considerable current interest   ~\cite{Simeoni2010,BrazhkinJPCB2011,BrazhkinPRE2012,
BrazhkinUFN2012,BrazhkinUFN2017,BellJCP2020}. 
The validity of SE relation is a strong indication of the liquid-like behaviour and hence should be related to the crossover.

\section{Results}

Here we analyse the available results for four simple systems with very diverse shape of pairwise interaction: Lenard-Jones, Coulomb (one-component plasma), screened Coulomb (Yukawa systems), and hard sphere system.

\subsection{Lennard-Jones liquids}     

The Lennard-Jones (LJ) potential is 
\begin{equation}
\phi(r)=4\epsilon\left[\left(\frac{\sigma}{r}\right)^{12}-\left(\frac{\sigma}{r}\right)^{6}\right], 
\end{equation}
where  $\epsilon$ and $\sigma$ are the energy and length scales (or LJ units), respectively. The reduced density and temperature expressed in LJ units are therefore $\rho_*=\rho\sigma^3$, $T_*=T/\epsilon$. The LJ system is one of the most popular and extensively studied model systems in condensed matter, because it combines relative simplicity with adequate approximation of interatomic interactions in real substances (e.g. liquified and solidified noble gases). 

Transport properties of LJ systems have been extensively studied in the literature. For recent reviews of available simulation data see e.g. Refs.~\cite{BellJPCB2019,HarrisJCP2020,AllersJCP2020}. Particularly extensive data sets on the viscosity and self-diffusion coefficients have been published by Meier {\it et al.}~\cite{Meier2002,MeierJCP_1,MeierJCP_2} and by Baidakov {\it et al.}~\cite{BaidakovFPE2011,BaidakovJCP2012}.   
These authors tabulated the transport data along different isotherms in a wide regions of the LJ system phase diagram. Though simulations protocols were different, the two datasets are in good agreement where they overlap~\cite{HarrisJCP2020}. 

\begin{figure}
\includegraphics[width=8.5cm]{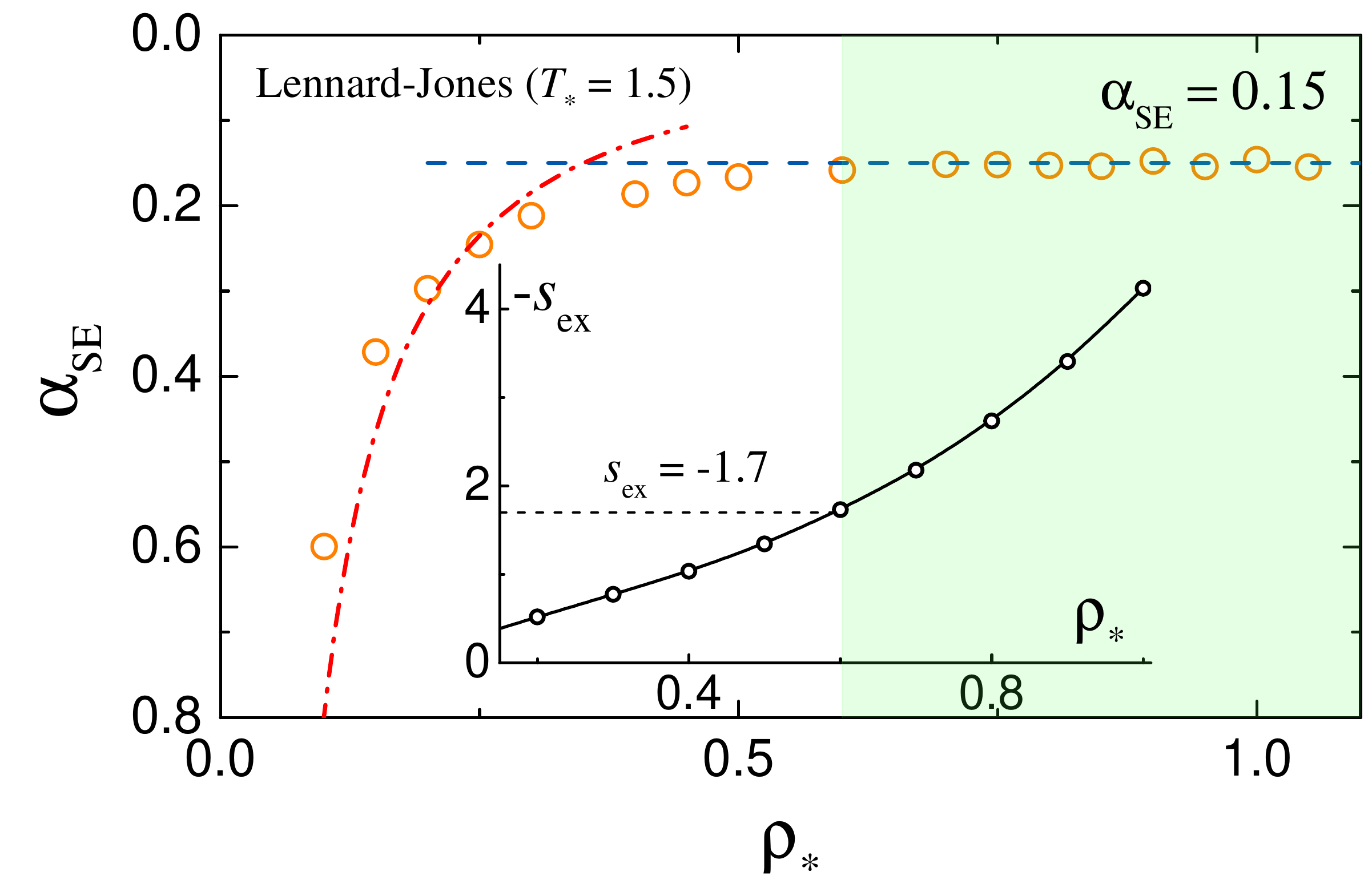}
\caption{(Color online) Stokes-Einstein parameter $\alpha_{\rm SE}$ versus the reduced density $\rho_*$ for a LJ liquid. The symbols correspond to MD simulation results from Ref.~\cite{Meier2002}. The dashed line is the fluid asymptote $\alpha_{\rm SE}\simeq 0.15$. The dash-dotted curve corresponds to the dilute HS gas asymptote $\alpha_{\rm SE}\simeq 0.037/\rho_{*}^{4/3}$. In the shaded area at $\rho_{*}> 0.6$ the SE coefficient is constant (lies in the narrow range $\alpha_{\rm SE}\simeq 0.15\pm 0.01)$. The inset shows the dependence of the minus excess entropy on the reduced density. Onset of the validity of the SE relation corresponds to $s_{\rm ex}\lesssim -1.7$. }
\label{Fig1}
\end{figure}

We have recently demonstrated that properly reduced transport coefficients (self-diffusion, shear viscosity, and thermal conductivity) of dense Lennard-Jones fluids along isotherms exhibit quasi-universal scaling on the density divided by its value at the freezing point, $\rho_{\rm fr}$~\cite{KhrapakPRE04_2021}. This implies that it is sufficient to consider a single isotherm. We chose the isotherm $T_*=1.5$ and employ the diffusion and viscosity coefficients tabulated in Ref.~\cite{Meier2002}. The resulting dependence of $\alpha_{\rm SE}$ on reduced density $\rho_*$ is plotted in Fig.~\ref{Fig1} (note the reversed vertical axis in the figure to highlight the level of accuracy of SE relation). It is observed that the SE coefficient drops with increasing density until it reaches the asymptotic value of $\alpha_{\rm SE}\simeq 0.15$. This is where the SE relation is satisfied. From pragmatical point of view we define the region of validity of SE relation as a region where the SE coefficient is located in a narrow range $\alpha_{\rm SE}\simeq 0.15\pm 0.01$. This occurs at $\rho_{*}\gtrsim 0.6$ and the corresponding region is shaded in Fig.~\ref{Fig1}. 

We may now ask whether the value $\simeq 0.15$ at which $\alpha_{\rm SE}$ saturates is consistent with Eq.~(\ref{DZ1}) above. The sound velocities of LJ liquids near the liquid-solid phase transition have been recently evaluated~\cite{KhrapakMolecules2020,KhrapakMolecules2021}.
Expressed in units of thermal velocity $v_{\rm T}=\sqrt{T/m}$ they turn out to be $c_l/v_{\rm T}\simeq 11.5$ and $c_t/v_{\rm T}\simeq 6$. Substituting this into Eq.~(\ref{DZ1}) we obtain $\alpha_{\rm SE}\simeq 0.15$, in excellent agreement with the results from MD simulations.  

In the low density regime we may use Eq.~(\ref{HStransport}) to estimate the diffusion and viscosity coefficients. For the SE relations this yields
\begin{equation}\label{HS2}
D\eta\left(\frac{\Delta}{T}\right)=\frac{15}{128\pi\rho_*^{4/3}}\simeq \frac{0.037}{\rho_*^{4/3}}.
 \end{equation} 
This density scaling is of course only approximate for dilute LJ gases. The actual transport cross sections are different from the hard-sphere model and specifics of scattering in the LJ potential has to be properly accounted for (see e.g. Refs.~\cite{HirschfelderBook,HirschfelderJCP1948,SmithJCP1964,
KhrapakPRE2014_scattering,KhrapakEPJD2014,Kristiansen2020} and references therein for some related works). Nevertheless, simple Eq.~(\ref{HS2}) already provides a reasonable approximation for MD data as documented in Fig.~\ref{Fig1} and reported previously in Ref.~\cite{KhrapakPRE04_2021}.      

The inset in Fig.~\ref{Fig1} shows the dependence of the minus reduced excess entropy $-s_{\rm ex}$ on the density as tabulated in Ref.~\cite{JaksePRE2003} for the LJ liquid isotherm $T_*=1.5$. The onset of validity of the SE relation corresponds to $s_{\rm ex}\lesssim -1.7$, according to a pragmatic definition given above. This approximate condition to be compared with the onset condition in other simple systems.  

The low-density asymptote $\alpha_{\rm SE}\simeq 0.037/\rho_*^{4/3}$ and the high-density asymptote $\alpha_{\rm SE}\simeq 0.15$ are intersecting at about $\rho_*\simeq 0.35$. This intersection can serve as a practical condition to locate the crossover between the gas-like and liquid-like regions on the LJ system phase diagram~\cite{KhrapakPRE04_2021}. At higher densities $\rho_*\gtrsim 0.6$ a fully liquid-like dynamics and transport emerge. According to the inset in Fig.~\ref{Fig1} the intersection of two asymptotes occurs at $s_{\rm ex}\simeq -0.9$.

Recently, diffusion and shear viscosity data of LJ fluid along nine supercritical isochores were analyzed with respect to the SE relation~\cite{CostigliolaJCP2019}.  It was shown that SE relation breaks down gradually at sufficiently high temperatures. This observation was rationalized in terms of the fact that properly 
reduced transport coefficient are approximately constant along the system's lines of constant excess entropy
(the isomorphs). This results in a quasi-universal dependence $D\eta(\Delta/T)= F(T/T_{\rm Ref}(\rho))$, where $F$ is some function and $T_{\rm Ref}(\rho)$ is the
temperature as a function of the density along a reference isomorph (the latter can be, but not necessarily, chosen as a freezing temperature~\cite{CostigliolaJCP2019}). These observations are correlating with ours: There is a curve on the phase diagram determined either by an equation $T=T(\rho)$ or $\rho=\rho(T)$, which is located above the freezing curve and is quasi-parallel to it in the first approximation. The SE relation is valid between these two curves.    


\begin{figure}
\includegraphics[width=8.0cm]{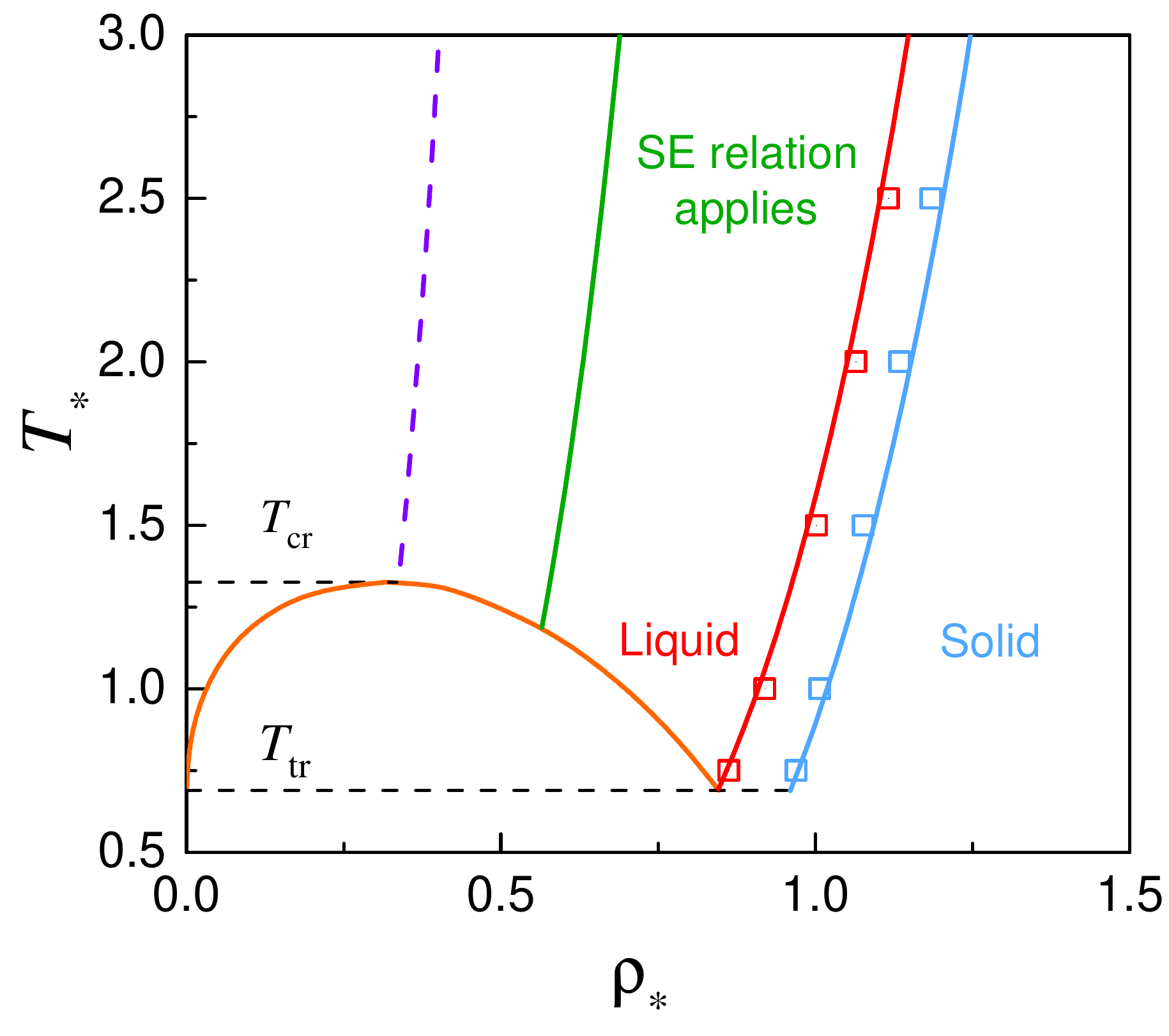}
\caption{(Color online) Different dynamical regimes on the LJ system phase diagram. The squares correspond to the fluid-solid coexistence boundaries as tabulated in Ref.~\cite{SousaJCP2012}; the corresponding curves are simple fits proposed in Ref.~\cite{KhrapakAIPAdv2016}. The liquid-vapour boundary is plotted using the formulas provided in Ref.~\cite{HeyesJCP2019}. The reduced triple point and critical temperatures are $T_{\rm tr}\simeq 0.694$~\cite{SousaJCP2012} and $T_{\rm cr}\simeq 1.326$~\cite{HeyesJCP2019}. Two additional curves appear in the phase diagram. The solid curve starting at a supercritical density corresponds to the condition $\rho/\rho_{\rm fr}=0.6$. The SE relation applies to the right from this curve. The dashed curve starting in he vicinity of the critical point corresponds to the condition $\rho/\rho_{\rm fr}=0.35$, where the dilute gaseous and dense liquid asymptotes for $\alpha_{\rm SE}$ intersect. This can be considered as a practical demarcation condition between the     gas-like and liquid-like regimes of atomic dynamics in supercritical fluids.}
\label{LJ_PD}
\end{figure}

The location of different regions on the LJ system phase diagram is sketched in Fig.~\ref{LJ_PD}. The phase boundaries are plotted using the results from Refs.~\cite{SousaJCP2012,KhrapakAIPAdv2016,HeyesJCP2019} (for details see the figure caption). Two additional curves are plotted in the phase diagram. The solid curve starting at a supercritical density corresponds to the condition $\rho/\rho_{\rm fr}=0.6$. For higher densities (to the right from this curve) the SE coefficient $\alpha_{\rm SE}$ becomes practically constant. This region has been identified as a region of validity of SE relation. The dashed curve starting in he vicinity of the critical point corresponds to the condition $\rho/\rho_{\rm fr}=0.35$, where the dilute gaseous and dense liquid asymptotes for $\alpha_{\rm SE}$ intersect (see Fig.~\ref{Fig1}). This can be considered as a practical demarcation condition between the     gas-like and liquid-like regimes of atomic dynamics in supercritical fluids. In terms of the excess entropy, the curve location roughly corresponds to $s_{\rm ex}\simeq -1$. It should be noted that another simple definition of a line separating gas-like and liquid-like fluid behaviors based on the properties of shear viscosity alone was proposed recently~\cite{BellJCP2020}. The separation line was defined by the location of the minimum of the macroscopically scaled shear viscosity coefficient when plotted as a function of the excess entropy. It was demonstrated that for hard sphere, Lennard-Jones, and inverse-power-law fluids, such a line is located at an excess entropy approximately equal to $s_{\rm ex} \simeq -2/3$. These definitions are not equivalent, but conceptually similar. The minimum of the reduced shear viscosity coefficient corresponds to the transition between dilute gaseous and dense liquid asymptotes for the transport properties. Not surprisingly, close value of excess entropy on the separation line are obtained from these two conditions.

In the region between the dashed and solid curves the liquid-like picture of atomic dynamics and transport processes is approached, but is not yet fully developed. Non-negligible deviations from the SE relation can be observed (see Fig.~\ref{Fig1}). The actual values of $\alpha_{\rm SE}$ are exceeding the asymptotic value $0.15$. Note that this can happen even in a liquid phase at slightly subcritical temperatures (see Fig.~\ref{LJ_PD}).        

Ohtori {\it et al}. investigated the validity of the SE relation in a wide region of the phase diagram of pure LJ liquids~\cite{OhtoriPRE2015,OhtoriPRE2017}. They concluded that the origin of the breakdown in the SE relation
can be traceable to the onset of the gaseous behaviour of the shear viscosity coefficient. The SE coefficient they reported is $\alpha_{\rm SE}\simeq 1/2\pi\simeq 0.16$ is however slightly larger than in other studies. The exact reason of this (small) disagreement is not yet identified. It should be noted that in Refs.~\cite{OhtoriPRE2015,OhtoriPRE2017} an explicit procedure to correct the self-diffusion coefficient for the effect of the finite system size in the periodic boundary conditions was implemented~\cite{DunwegJCP1993,YehJPCB2004} (A particularly simple form of this correction $D_{\infty}\simeq D_{N}(1+N^{-1/3})$ has been suggested recently~\cite{KhrapakMolPhys2019}; here $D_{\infty}$ is the infinite-size system diffusion coefficient and
$D_N$ is the diffusion coefficient evaluated for $N$ particles in a cubic cell with periodic boundary conditions). This results in a slightly larger diffusion coefficient in the thermodynamic limit compared to the actual value from MD simulations and can potentially be responsible for some inconsistency.    

\subsection{One-component plasma}

The one-component plasma (OCP) model is an idealized
system of point charges immersed in a neutralizing 
uniform background of opposite charge (e.g. ions in the immobile background of electrons or vice versa)~\cite{BrushJCP1966,deWitt1978,BausPR1980,IchimaruRMP1982,
StringfellowPRA1990}. This model is of considerable practical interest  in the plasma related context. From the fundamental point of view OCP is characterized by a very soft and long-ranged Coulomb interaction potential, 
\begin{equation}
\phi(r)= e^2/r, 
\end{equation}
where $e$ is the electric charge. This potential is very much softer than the Lennad-Jones potential considered above. This makes OCP a very important model reference system to verify the validity of various theories and approaches to soft condensed mater. 

The particle-particle correlations and thermodynamics of the OCP are characterized by a single dimensionless coupling parameter $\Gamma=e^2/aT$, where $a$ is the Wigner-Seitz radius. The coupling parameter essentially plays the role of an inverse temperature or inverse interatomic separation. In the limit of weak coupling (high temperature, low density), $\Gamma\ll 1$, the OCP is in a disordered gas-like state. Correlations increase with coupling and, at $\Gamma\gtrsim 1$, the OCP exhibits properties characteristic of a fluid-like phase (low temperature, high density). The fluid-solid phase transition occurs at $\Gamma\simeq 174$~\cite{IchimaruRMP1982,DubinRMP1999,KhrapakCPP2016}.     

\begin{figure}
\includegraphics[width=8.5cm]{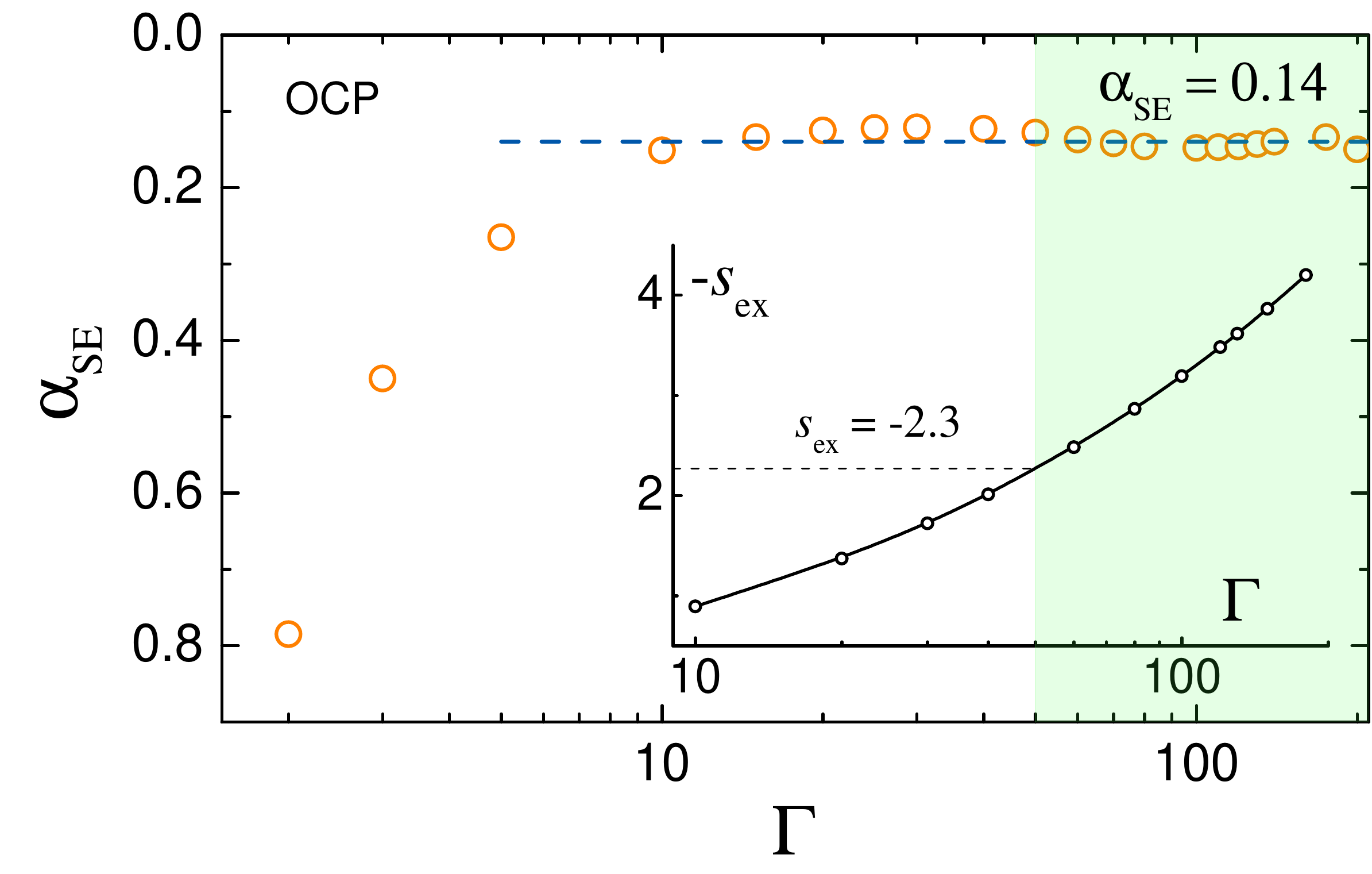}
\caption{(Color online) Stokes-Einstein parameter $\alpha_{\rm SE}$ versus the coupling parameter $\Gamma$ for a OCP fluid. The symbols correspond to MD simulation results from Refs.~\cite{DaligaultPRE2012,DaligaultPRE2014}. The dashed line is a strong coupling asymptote $\alpha_{\rm SE}\simeq 0.14$. In the shaded area at $\Gamma\gtrsim 50$ the SE coefficient is practically constant (lies in a narrow range $\alpha_{\rm SE}\simeq 0.14\pm 0.01$). The inset shows the dependence of the minus excess entropy on the coupling parameter. Onset of the validity of the SE relation corresponds to $s_{\rm ex}\lesssim -2.3$. }
\label{Fig2}
\end{figure}

Transport properties of the OCP and related system are very well investigated in classical MD simulations. Extensive data on the self-diffusion~\cite{DaligaultPRL2006,DaligaultPRL2012,DaligaultPRE2012,
KhrapakPoP2013} and shear viscosity~\cite{DonkoPoP2000,SalinPRL2002,BasteaPRE2005,
DaligaultPRE2014,KhrapakPRE01_2021} have been published and discussed in the literature. Here we have used an accurate fitting formula for the self-diffusion coefficient proposed in Ref.~\cite{DaligaultPRE2012} along with the MD data on the shear viscosity coefficient tabulated in Ref.~\cite{DaligaultPRE2014}. The resulting dependence of the SE coefficient $\alpha_{\rm SE}$ on the coupling parameter $\Gamma$ is plotted in Fig.~\ref{Fig2}. 

We observe that the slope of the dependence $\alpha_{\rm SE}$ on $\Gamma$ changes at approximately $\Gamma\simeq 10$. The strong coupling asymptote is $\alpha_{\rm SE}\simeq 0.14$. The weak coupling asymptote is not shown in the Figure because the conventional Coulomb scattering theory is only applicable at $\Gamma\ll 1$. At $\Gamma\gtrsim 50$ the SE coefficient lies in a narrow range $\alpha_{\rm SE}\simeq 0.14\pm 0.01$. This is where we identify the SE relation to be valid (shaded area in Fig.~\ref{Fig2}), although already from $\Gamma\simeq 10$ the deviations from the strong coupling asymptote are relatively small. 

In the OCP fluid the longitudinal mode does not exhibit the acoustic-like dispersion, it is a plasmon mode  (for an example of numerically computed and analytical dispersion relations see e.g. Refs.~\cite{GoldenPoP2000,SchmidtPRE1997,KhrapakAIPAdv2017,
KhrapakIEEE2018}). Formally, this corresponds to the condition $c_t/c_l=0$ and one could expect $\alpha_{\rm SE}\simeq 0.13$ in he strong coupling limit. However, the acoustic Debye spectrum used in Zwanzig's derivation is itself not valid in the OCP case. Apparently, the longitudinal mode provides non-negligible contribution and this can explain the difference of actual $\alpha_{\rm SE}$ from its minimal limit. 

The inset in Fig.~\ref{Fig2} shows the dependence of the minus reduced excess entropy $-s_{\rm ex}$ on the coupling parameter $\Gamma$ as tabulated in Ref.~\cite{LairdPRA1992}. The change in the slopes of asymtotes at $\Gamma\simeq 10$ corresponds to $s_{\rm ex}\simeq -0.9$. The onset of validity of the SE relation at $\Gamma\simeq 50$ corresponds to $s_{\rm ex}\lesssim -2.3$.


\subsection{Screened Coulomb (Yukawa) fluids}

The Yukawa systems represent a collection of point-like charges immersed into a neutralizing polarizable medium (usually conventional electron-ion plasma), which provides screening. The pairwise Yukawa repulsive interaction  potential (also known as screened Coulomb or Debye-H\"uckel potential) is
\begin{equation}\label{Yukawa}
\phi(r)=(e^2/r)\exp(-\kappa r/a),
\end{equation}
where $\kappa$ is the dimensionless screening parameter, which is the ratio of the Wigner-Seitz radius to the plasma screening length. Yukawa potential is widely used as a reasonable first approximation for actual  interactions in three-dimensional isotropic complex plasmas and colloidal suspensions~\cite{TsytovichUFN1997,FortovUFN,FortovPR,
IvlevBook,KhrapakCPP2009,KhrapakPRL2008,KlumovUFN2010,ChaudhuriSM2011,LampePoP2015}. 

The dynamics and thermodynamics of Yukawa systems are  characterized by two reduced parameters, $\Gamma$ and $\kappa$. Detailed phase diagrams of Yukawa systems are available in the literature~\cite{RobbinsJCP1988,HamaguchiJCP1996,HamaguchiPRE1997,
VaulinaJETP2000,VaulinaPRE2002}. 
Note that the screening parameter $\kappa$ determines the softness of the interparticle repulsion. It varies from the very soft and long-ranged Coulomb potential at $\kappa\rightarrow 0$ (corresponding to the OCP limit considered above) to the hard-sphere-like interaction limit at $\kappa\rightarrow \infty$. In the context of complex plasmas and colloidal suspensions the relatively ``soft'' regime, $\kappa\sim {\mathcal O}(1)$, is of particular interest. 

\begin{figure}
\includegraphics[width=8.5cm]{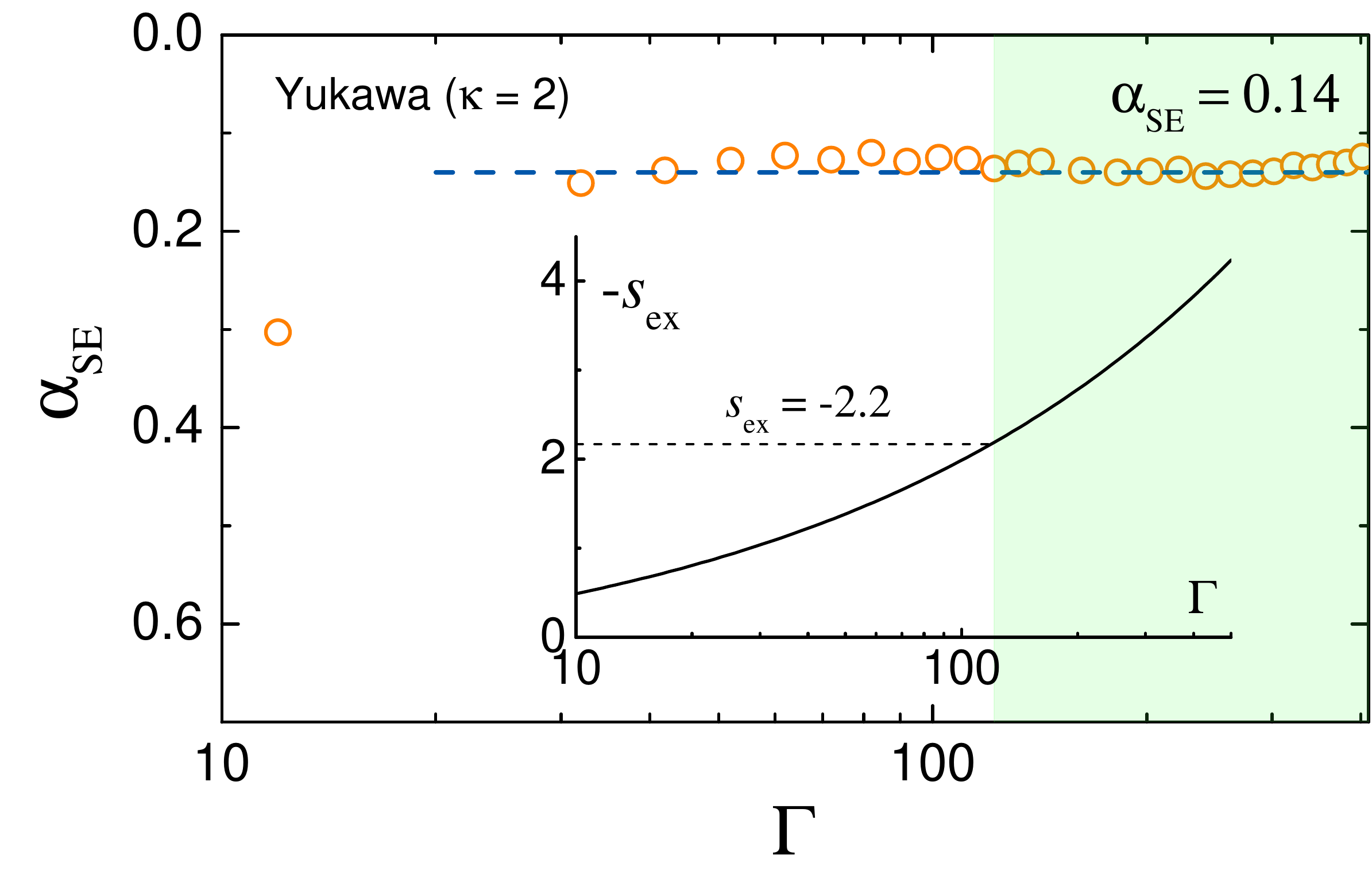}
\caption{(Color online) Stokes-Einstein parameter $\alpha_{\rm SE}$ versus the coupling parameter $\Gamma$ for a Yukawa fluid with $\kappa=2$. The symbols correspond to MD simulation results from Refs.~\cite{DaligaultPRE2012,DaligaultPRE2014}. The dashed line is the fluid asymptote $\alpha_{\rm SE}\simeq 0.14$. In the shaded area at $\Gamma\gtrsim 120$ the SE coefficient is practically constant (lies in a narrow range $\alpha_{\rm SE}\simeq 0.14\pm 0.01$). The inset shows the dependence of the minus excess entropy on the coupling parameter. Onset of the validity of the SE relation corresponds to $s_{\rm ex}\lesssim -2.2$. }
\label{Fig3}
\end{figure}

The phenomena of self-diffusion and shear viscosity in three-dimensional Yukawa fluids have been relatively well investigated and understood~\cite{RobbinsJCP1988,OhtaPoP2000,SanbonmatsuPRL2001,
SaigoPoP2002,VaulinaPRE2002,SalinPRL2002,SalinPoP2003,FaussurierPRE2003,
DonkoPRE2008,DaligaultPRE2012,KhrapakPoP2012,DaligaultPRE2014,
KhrapakJPCO2018,KhrapakAIPAdv2018,
KahlertPPR2020} (there have been also considerable interest to two-dimensional systems related to laboratory realizations of dusty plasmas, which are not considered here). For our present purposes we combine the accurate fits for the self-diffusion coefficient in Yukawa fluids from Ref.~\cite{DaligaultPRE2012} with numerical data on shear viscosity coefficient tabulated in Ref.~\cite{DaligaultPRE2014} for $\kappa = 2$. The results are plotted in Fig.~\ref{Fig3}. 

The picture emerging from Fig.~\ref{Fig3} is similar to that in the OCP case, except higher $\Gamma$ values are involved. The slope of the dependence $\alpha_{\rm SE}$ on $\Gamma$ changes at approximately $\Gamma\simeq 30$. The strong coupling asymptote is $\alpha_{\rm SE}\simeq 0.14$, same as in the OCP case. The weak coupling asymptote is not shown for the same reason as for the OCP. At $\Gamma\gtrsim 120$ the SE coefficient lies in a narrow range $\alpha_{\rm SE}\simeq 0.14\pm 0.01$ and as previously we identify this as the region of validity of the SE relation (shaded area in Fig.~\ref{Fig3}). As previously, the deviations from the asymptotic strong coupling value are already relatively small at $\Gamma\gtrsim 30$. 

The inset in Fig.~\ref{Fig2} shows the dependence of the minus reduced excess entropy $-s_{\rm ex}$ on the coupling parameter $\Gamma$. The curve is calculated using the Rosenfeld-Tarazona scaling of the thermal component of the excess internal energy~\cite{RosenfeldMolPhys1998}. This scaling has been proven to be very useful in constructing practical models for the thermodynamic of Yukawa fluids~\cite{KhrapakPRE02_2015,KhrapakJCP2015,ToliasPoP2019}. The particular form used here is taken from Ref.~\cite{RosenfeldPRE2000}:    
\begin{displaymath}
s_{\rm ex}\simeq -4.5\left(\frac{\Gamma}{\Gamma_{\rm fr}}\right)^{2/5}+0.5,
\end{displaymath}  
where it is assumed that the excess entropy at freezing is $s_{\rm ex}\simeq -4$. Here $\Gamma_{\rm fr}$ denotes the value of the coupling parameter at the fluid-solid phase transition, which is $\Gamma_{\rm fr}\simeq 440$ at $\kappa=2$~\cite{HamaguchiPRE1997}. The change in the slopes of asymptotes at $\Gamma\simeq 30$ corresponds to $s_{\rm ex}\simeq -1.0$. The onset of validity of the SE relation at $\Gamma\simeq 120$ corresponds to $s_{\rm ex}\lesssim -2.2$.

\subsection{Hard-Sphere fluids}

The fourth system considered is the fluid consisting of hard spheres. The HS interaction potential is extremely hard and short ranged. The interaction energy is infinite for $r<\sigma$ and is zero otherwise, where $\sigma$ is the sphere diameter. The HS system is a very important simple model for the behaviour of condensed matter in its various states~\cite{SmirnovUFN1982,MuleroBook,
PuseyPhylTrans2009,ParisiRMP2010,BerthierRMP2011,
KlumovPRB2011,DyreJPCM2016}.  

\begin{figure}
\includegraphics[width=8.5cm]{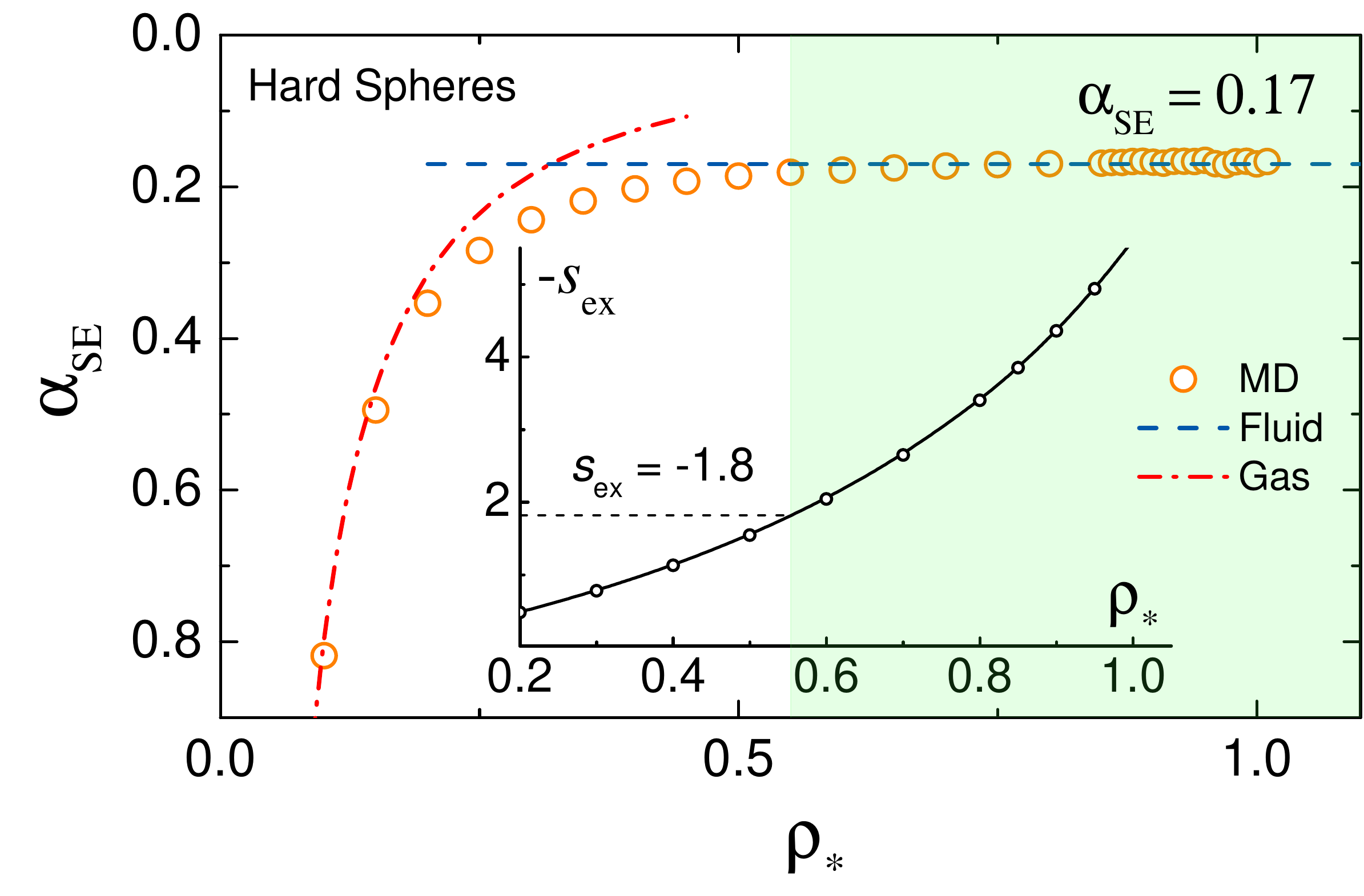}
\caption{(Color online) Stokes-Einstein parameter $\alpha_{\rm SE}$ versus the reduced density $\rho_*$ for a HS fluid. The symbols correspond to MD simulation results from Refs.~\cite{Pieprzyk2019,Pieprzyk2020}. The dashed line is the dense fluid asymptote $\alpha_{\rm SE}\simeq 0.17$. The dash-dotted curve corresponds to the dilute HS gas asymptote $\alpha_{\rm SE}\simeq 0.037/\rho_{*}^{4/3}$. In the shaded area at $\rho_{*}> 0.55$ the SE coefficient is constant (lies in the narrow range $\alpha_{\rm SE}\simeq 0.17\pm 0.01)$. The inset shows the dependence of the minus excess entropy on the reduced density. Onset of the validity of the SE relation corresponds to $s_{\rm ex}\lesssim -1.8$.}
\label{Fig4}
\end{figure}

In HS systems the thermodynamic and transport properties  depend on a single reduced density parameter $\rho_*=\rho\sigma^3$ (the packing fraction, $\pi\rho\sigma^3/6$, is also often used). Transport properties of HS fluids have been extensively studied (see e.g. Ref.~\cite{MuleroBook} for a review). For our present purpose we have used the recent MD simulation results by Pieprzyk {\it et al}.~\cite{Pieprzyk2019,Pieprzyk2020}. The use of large simulation systems and long simulation times allowed accurate prediction of the self-diffusion and shear viscosity coefficients in the thermodynamic limit. Based on the tabulated data we have evaluated the SE coefficient and plotted it as a function of the reduced density in Fig.~\ref{Fig4}.

In Fig.~\ref{Fig4} we observe that the data points stick to the two asymptotes: the gaseous Eq.~(\ref{HS2}) at low densities and the liquid-like $\alpha_{\rm SE}\simeq 0.17$ at sufficiently high density. The inset shows the dependence of the minus excess entropy on the reduced density. The asymptotes are intersecting at $\rho_*\simeq 0.32$, which corresponds to $s_{\rm ex}\simeq -0.8$. The shaded region in Fig.~\ref{Fig4} is where the SE coefficient lies in a narrow range $\alpha_{\rm SE}\simeq 0.17\pm 0.01$. This is approximately the regime of SE relation validity according to our pragmatic definition.
Numerically, the onset of SE relation validity  occurs at $\rho_*\gtrsim 0.55$, which corresponds to 
$s_{\rm ex}\simeq -1.8$.

It should be pointed out that Zwanzig's derivation of the SE relation is obviously inconsistent with the dynamical picture in HS fluids. In contrast to softer interactions, the velocity autocorrelation function $Z(t)$ rapidly vanishes after the first rebound against the initial cage and does not exhibit a pronounced oscillatory character assumed in Zwanzig's derivation~\cite{AlderPRL1967,WilliamsPRL2006,Daligault2020}. 
Nevertheless, we see that the SE relation is still satisfied even in this case. Dense HS fluids support the acoustic-like longitudinal and transverse collective modes~\cite{BrykJCP2017} (although with a forbidden long wavelength region for the transverse mode, the so-called  ''$k$-gap''~\cite{BrykJCP2017,YangPRL2017,KryuchkovSciRep2019,KhrapakPRE05_2021}.
We can compare the SE coefficients obtained in numerical simulations with that formally emerging from Eq.~(\ref{DZ1}). The ratio of the transverse to longitudinal velocities tends to $c_t/c_l\simeq 0.5$ on approaching the HS limit~\cite{KhrapakPRE05_2021}. This yields a theoretical estimate $\alpha_{\rm SE}\simeq 0.15$, somewhat smaller than the actual value obtained from MD simulation. The difference can be possibly related to the inconsistencies between theoretical assumptions and actual dynamical picture in HS fluids.     

\section{Conclusion}

The applicability of the Stokes-Einstein relation without the hydrodynamic radius to dense simple fluids has been investigated in detail. Four model systems with very different pairwise interaction potential have been considered: Lennard-Jones, one-component plasma, screened Coulomb (Yukawa) and hard sphere fluids. From the presented evidence the following main conclusions can be made.

In all systems considered the SE relation understood as a constancy of the product $D\eta(\Delta/T)=\alpha_{\rm SE}={\rm const}$, is satisfied to a good accuracy in some region of the phase diagram adjacent to the freezing curve. The value of the SE coefficient correlates with the softness of the interaction potential (via the ratio of the longitudinal and transverse sound velocities). For soft long-ranged interactions such as in one-component plasma and Yukawa fluids it is smaller than for the more steep Lennard-Jones and HS fluids. This is consistent with the Zwanzig's construction because softer potentials usually result in smaller ratios of transverse-to-longitudinal sound velocities. For the HS model the actual $\alpha_{\rm SE}$ obtained from simulations is somewhat higher than theoretically expected. However, for the HS fluid the vibrational picture of atomic dynamics assumed in theory is violated and some discrepancies should not be very surprising.       

The relative width of the density region where the SE relation is valid varies considerably. For the LJ and HS systems the onset of validity can be estimated as $\rho/\rho_{\rm fr}\simeq 0.6$. For the OCP and Yukawa systems the onset of validity is roughly at $\Gamma/\Gamma_{\rm fr}\sim 0.3$, which corresponds to $\rho/\rho_{\rm fr}\sim 0.03$ (since $\Gamma\propto \rho^{1/3}$). Thus, density divided by the density at the freezing point is not a reliable measure of SE relation validity. In this respect, the excess entropy appears as a much more convenient indicator of its validity. In all cases considered, the onset point roughly corresponds to $s_{\rm ex}\simeq -2.0\pm 0.3$. It remains valid up to the liquid-solid phase transitions, which is characterised by $s_{\rm ex}\simeq -4$ for the systems considered.                   

Another important observation is that there exist two clear asymptotes for the product $D\eta(\Delta/T)$. Near the freezing point it approaches a slightly system-dependent constant value. Far away from the freezing point, in a disordered gas phase, this product decreases towards the more ordered fluid state (i.e. with density). The intersection of these two asymptotes is characterised by almost identical values of excess entropy, $s_{\rm ex}\simeq -0.9 \pm 0.1$. This can be used as a convenient practical condition for the crossover between the gas-like and liquid-like regions on the phase diagram. 

In future studies it would be interesting to ascertain to which extent this picture applies to real fluids, mixtures and other types of interaction potential, such as for instance bounded potentials, which are of considerable impact in soft matter research.   


\bibliography{SE_Ref}

\end{document}